\documentclass[twocolumn,prd,showpacs,aps]{revtex4}
\usepackage{graphicx}% Include figure files
\usepackage{dcolumn}% Align table columns on decimal point

\newcommand{\na}{\nabla}
\newcommand{\ep}{\epsilon}

\newcommand{\vphi}{\varphi}
\newcommand{\lraw}{\longrightarrow}

\newcommand{\pa}{\partial}
\newcommand{\td}{\tilde}

\newcommand{\Sla}[1]{\slash\!\!\!\! #1}
\newcommand{\beq}[1]{\begin{eqnarray}\label{#1}}
\newcommand{\eeq}{\end{eqnarray}}
\newcommand{\net}{{\cal N}=1}
\newcommand{\gym}{g_{_{\rm YM}}}
\newcommand{\llraw}{-\!-\!\!\!\lraw}
\renewcommand{\footnote}{\footnote}

\begin{document}

\title{Green Functions of ${\cal N}=1$ SYM and Radial/Energy-Scale Relation}

\author{Xiao-Jun Wang}

\email{wangxj@itp.ac.cn} \affiliation{Institute of Theoretical
Physics, Beijing, 100080, P.R.China}

\author{Sen Hu}
\email{shu@ustc.edu.cn} \affiliation{Department of Mathematics,
University of Science and
Technology of China, \\
AnHui, HeFei, 230026, P.R. China}

\begin{abstract}
We study counter-terms of one- and two-point Green functions of
some special operators in ${\cal N}=1$ SYM from their SUGRA duals
from the consideration of AdS/CFT or gauge/gravity correspondence.
We consider both the Maldacena-Nunez solution and the
Klebanov-Strassler-Tseytlin solution that are proposed to be SUGRA
duals of ${\cal N}=1$ SYM. We obtain radial/energy-scale relation
for each solution by comparing SUGRA calculations with the field
theory results. Using these relations we evaluate the
$\beta$-function of ${\cal N}=1$ SYM. We find that the leading
order term can be accurately obtained for both solutions and the
higher order terms exhibit some ambiguities. We discuss the origin
of these ambiguities and conclude that more studies are needed to
check whether these SUGRA solutions are exactly dual to ${\cal
N}=1$ SYM.

\end{abstract}


\maketitle

\section{Introduction}

Due to the remarkable success of AdS/CFT
correspondence~\cite{Mald98} more and more attention has been
attracted to the study of gauge/gravity duality with less
supersymmetry and the gauge theory part is not conformally
invariant. The $d=4,\;\net$ super Yang-Mills theory (SYM) turned
out to be extremely interesting with rich results. Nonsingular
supergravity (SUGRA) solutions are constructed and proposed to be
dual of the $\net$ super Yang-Mills theory. With these solutions
we are able to get information of the dual quantum gauge theory at
all energy scales. So far two SUGRA backgrounds were found:
\begin{itemize}
\item The Maldacena-Nu$\td{\rm n}$ez (MN) solution\cite{MN01}:
It corresponds to a large ($N$) number of D5-branes wrapped on a
supersymmetric two-cycle inside a Calabi-Yau threefold. It is dual
to pure $\net$ $SU(N)$ SYM.
\item The Klebanov-Strassler-Tseytlin (KST)
solution~\cite{KT00,KS00}: It describes the geometry of the warped
deformed conifold when one places M D3-branes and N fractional
D3-branes at the apex of the conifold. It is dual to a certain
$\net$ supersymmetric $SU(N+M)\times SU(M)$ gauge theory. If $M$
is a multiple of $N$, then this theory flows to $SU(N)$ in the IR,
via a chain of duality cascade which reduces the rank of the gauge
group by $N$ units at each cascade jump. Thus at the end of the
duality cascade the gauge theory is effectively pure $\net$
$SU(N)$ SYM.
\end{itemize}

The verification on the above gauge/gravity duality is to perform
calculations for correlation functions in terms of the SUGRA
duals, in particular for Green functions in the non-conformal
case. Since the KST solution is asymptotically AdS$_5\times X^5$
in the UV, where $X^5$ is an Einstein manifold, the extension to
KST solution is direct\cite{Kras02}. That is the classical action
of SUGRA is identified with the generating function of ${\cal
N}=1$ SYM at the large $N$ limit and bulk field/boundary operator
correspondence is established via solving the SUGRA equations of
motion of bulk field $\phi(x,r)$ subject to boundary conditions in
the UV, i.e., for $r\to\infty$, $\phi(x,r)=r^\Delta\phi_0(x)$
where $r$ is radial of the asymptotic AdS space, and $\Delta$ is
related to (mass) dimension of the boundary operator. The
extension to MN solution, however, has to be re-examined since MN
solution is no longer asymptotic to AdS. MN background can be
treated as a domain wall solution of a truncated $d=7$ gauge
supergravity\cite{CLP00,DLM02} and this $d=7$ gauged supergravity
is obtained by compactification of IIB supergravity on $S^3$. Thus
the partition function of IIB superstring in the MN background can
naturally be identified with the generating function of ${\cal
N}=1$ SYM. At large $N$ limit we can identify the classical SUGRA
action in the MN background with the generating function of
${\cal N}=1$ SYM. The bulk field/boundary operator correspondence
is also established via similar method as in KST solution.
Because the transverse space of MN background is parameterized as
$S^2\times S^3$ respectively in UV, $\Delta$ is determined by two
parameters which are eigenvalues of the spherical harmonic
functions on $S^2$ and on $S^3$. Meanwhile $\Delta$ in KST
solution is determined by the mass $m$ of the bulk field in
AdS$_5$.

When we try to extend gauge/graivity duality beyond the
conformally invariant case with maximal supersymmetry a well-known
problem is that fluctuation equations on those complicated
background can not be strictly solved. It prevents us from
extracting enough information about the Green functions. In
general there are two kinds of useful information included in
Green functions from their SUGRA dual description. One is the
singular behavior of the Green function which can be used to
directly check the duality between SYM and their SUGRA dual. For
KST solution it was partially discussed in ref.\cite{Kras02}.
Another one is the structure of counter-terms of the Green
function from which we can obtain information of the
renormalization group. In Ref.\cite{WH02} we have shown that it
can be extracted even though we only got near-boundary asymptotic
solution of the fluctuation equation. In other words with those
asymptotic solution and the UV/IR relation\cite{PP98} we can
analyze the divergence structure of the Green function of SYM from
their SUGRA dual. In particular the relation between the radial in
SUGRA configuration and the energy scale in the dual gauge theory
can in principle be established via comparing counter-terms of the
Green functions calculated from SUGRA side and from SYM side
respectively. In Ref.\cite{WH02} the ${\cal N}=2$ case was
successfully studied. The purpose of the present paper is to study
the structure of counter-terms of the Green functions and the
consequent radial/energy-scale relation with $\net$ supersymmetry
in the framework of both MN and KST background.

The radial/energy-scale relation is crucial in the study of
gauge/gravity duality. For conformally invariant case this
relation has been well established\cite{Mald98,PP98, Relation}.
For non-conformally invariant theories it was suspected that to
establish such a relation may be difficult for the relation maybe
ambiguous. However at least for ${\cal N}=2$ case it was shown
that such a relation can be unambiguously obtained via several
different ways\cite{WH02,DLM02}. For $\net$ case although there
are some complications many interesting results have been
achieved: the authors of Ref.\cite{HKO01} extend
radial/energy-scale relation to KST solution and show that it is
right at the leading order. In Ref.\cite{ABCPZ01}, the leading
order expression of radial/energy-scale relation in MN solution
was also achieved via comparing $\beta$-functions calculated from
both the SUGRA side and the field theory side. In particular the
authors of Ref.\cite{DLM02} (DLM) suggested radial/energy-scale
relation in $\net$ case can be completely established via
considering the phenomena of gaugino condensation. In concrete
they considered a scalar SUGRA field whose boundary value couples
to a gaugino bi-linear form and they identified the scaling
behavior of this field to the inverse of gaugino condensation.
They have applied this idea to MN solution and obtain
radial/energy scale relation for the MN solution and consequently
obtain $\beta$-function for all loops. However in Ref.\cite{OS02}
it was shown that DLM's result is right at the leading order only
and a correction has to be made to compensate the effect of
gaugino condensation in order to get the right result for higher
order. Surprisingly DLM's idea has been applied to KST
solution\cite{Ime02} and the right $\beta$-function for all orders
was obtained.

In this paper, following the method shown in Ref.\cite{WH02}, we
calculate one- and two-point Green functions of certain operators
of SYM via their SUGRA dual descriptions. These results in general
involve a large radial cut-off. It corresponds to UV cut-off in
quantum field theory (QFT) calculation. Therefore we can
investigate the divergence behavior of Green functions resulted
from SUGRA and from QFT respectively and compare coefficients of
these terms possessing same pole behavior. Then we will obtain a
radial/energy-scale relation for this gauge/gravity
correspondence. Of course the ${\cal N}=1$ case is very different
from the ${\cal N}=2$ case. We will discuss those differences in
section~\ref{sec2} and in section~\ref{sec4}.

The paper is organized as follows. In section~\ref{sec2}  we
consider the bilinear gaugino operator. We calculate gaugino
condensation (one-point function) and two-point function of this
operator for MN solution. We then extract radial/energy-scale
relation for MN solution and calculate the $\beta$-function to
check this relation. In section~\ref{sec3} the same steps will be
performed for KST solution. However for KST solution we
conveniently consider dimension four operator instead of the
bilinear gaugino operator. In section~\ref{sec4} we discuss some
ambiguities when we try to exactly match results from SUGRA
description with the one from QFT. We also devote a brief summary
in this section.

\section{Green Functions and Radial/Energy-Scale Relation for
MN Solution}
\label{sec2}

\subsection{The MN Solution}

The MN solution can be obtained via finding a domain wall solution
of a seven dimension gauge supergravity after truncating its
$SO(4)$ gauge group to $SU(2)$\cite{DLM02}. This solution is
parameterized by three dimensionless scalar functions $k,\;g$, and
$a$. They are in general independent of the five transverse space
dimensions whose geometry is $S^3\times S^2$. Therefore we can
further compactify this truncated 7-d supergravity on $S^2$ and
get a 5-d effective action
 \beq{1}
   S=\eta\int d^4x d\rho &e^{2k}\{&4\pa_i k\pa^ik-2\pa_i g\pa^ig
     \nonumber \\  &&-\frac{\pa_i a\pa^ia}{2}e^{-2g}+V\},
 \eeq
where the metric is $g_{ij}=\eta_{ij}$, $\rho$ is a dimensionless
radial parameter, $\eta\sim \alpha'^{-2}$ and
 \beq{2}
   V=4+2e^{-2g}-\frac{(1-a^2)^2}{4}e^{-4g}.
 \eeq
The MN solution is obtained by assuming that functions $k,\;g$,
and $a$ depend on $\rho$ only. Explicitly we read off this
solution in the ten-dimensional string frame
metric~\cite{MN01,DLM02,CV97}
\begin{widetext}
 \beq{3}
   ds^2&=&e^\Phi\left[dx^2_{1,3}+\frac{e^{2g}}{\lambda^2}
      (d\theta^2+\sin^2{\theta}d\phi^2)
     +\frac{1}{\lambda^2}d\rho^2+\frac{1}{4\lambda^2}\sum_{a=1}^{3}
     (w^a-\frac{A^a}{\lambda})^2\right], \nonumber \\
   e^{2g}&=&\rho\coth{2\rho}-\frac{\rho^2}{\sinh^2{2\rho}}
             -\frac{1}{4},  \\
   e^{2\Phi}&=&e^{2k-2g}=ce^{-g}\sinh{2\rho}, \hspace{0.8in}
   a=\frac{2\rho}{\sinh{2\rho}}, \nonumber
 \eeq
\end{widetext}
where $\lambda^{-2}=Ng_s\alpha'$, $c$ are integral constants,
$w^a$ parameterizes the 3-sphere, and gauge field $A_a$ are
written as
 \beq{4}
   A^1=-\lambda ad\theta,\quad
   A^2=\lambda a\sin{\theta}d\phi,\quad
   A^3=-\lambda\cos{\theta}d\phi. \nonumber \\
 \eeq
The value of the Yang-Mills coupling is given in terms of the
volume of the sphere $S^{2}$ by\cite{DLM02}
 \beq{5}
   \frac{1}{\gym^2}=\frac{{\rm Vol}_{S^2}}{g^2_{D_6}}=
   \frac{N}{16\pi^2}Y(\rho)E\left(\sqrt{\frac{Y(\rho)-1}
   {Y(\rho)}}\right),
 \eeq
where the function $Y$ is defined as
 \beq{6}
  Y(\rho)=4e^{2g}+a^2=4\rho\coth{2\rho}-1,
 \eeq
and $E$ is a complete elliptic integral
 \beq{7}
   E(k)=\int_0^{\pi/2}dx\sqrt{1-k^2\sin^2{x}}.
 \eeq

\subsection{Green functions and radial/energy scale relation}

In the ${\cal N}=2$ case we studied Green functions of dimension
four operator, ${\cal O}_4(x)=Tr(F_{\mu\nu}F^{\mu\nu}+\cdots)$,
which couples to fluctuation of gauge coupling. However it is
difficult to consider similar operator for the MN solution. The
reason is that the gauge coupling $g^{-2}_{YM}$ is a complicated
nonlinear function of fields $g$ and $a$. Because of this it is
very hard to identify fluctuation of gauge coupling to one of the
dynamical fields $k,\;g$ or $a$. Consequently the fluctuation
equation corresponding to the above operator can not be derived.
Fortunately it is well-known that the field $a$ directly couples
to gaugino bilinear operator\cite{HKO01},
 \beq{8} {\cal O}_3(x)=Tr\bar{\psi}(x)\psi(x). \eeq
Therefore it is convenient to calculate Green functions of ${\cal
O}_3(x)$.

Let us consider the fluctuation of $a$ in background~(\ref{3}),
 \beq{9} a(x,\rho)\rightarrow \bar{a}(\rho)+\frac{\vphi(x,\rho)}
  {\lambda}, \eeq
where $\bar{a}(\rho)$ is a background solution in Eq.~(\ref{3}),
 \beq{10} \vphi(x,\rho)=s(\rho)\int\frac{d^4p}{(2\pi)^4}
   e^{ip\cdot x} \eeq
is the fluctuation normalized to $s(\rho\to\infty)=1$ at the
boundary. The field equation for this fluctuation reads off
 \beq{11} \ddot{s}&+&2(\dot{k}-\dot{g})\dot{s}-(n+\frac{p^2}{\lambda^2})s+
   e^{-2g}(m-3a^2)s \nonumber \\
   &+&2\bar{a}(\dot{\delta k}-\dot{\delta g})
   -\bar{a}(1-\bar{a}^2)e^{-2\delta g}=0, \eeq
where dot denotes differential on $\rho$, $\delta k$ and $\delta
g$ are fluctuations of fields $k$, and $g$ respectively and the
constants $m$ and $n$ are associated to eigenvalues of the
spherical harmonic functions of $S^2$ and $S^3$ respectively.

The one-point function of ${\cal O}_3$ or the gaugino condensation
depends on normalization of fluctuations yet it is independent of
the explicit expression of fluctuation. Taking cut-off
$\rho_0\to\infty$ at a region near the boundary we easily get
 \beq{12} <Tr\bar{\psi}\psi>=\frac{\eta}{\lambda}\left.e^{2k-2g}
 \dot{a}\right|_{\rho=\rho_0}=\frac{c\eta}{\lambda}\rho_0^{1/2}
   +O(\rho_0^{-1/2})
 \eeq

The gaugino condensation of ${\cal N}=1$ SYM has been computed in
terms of quantum field theory methods in different
schemes~\cite{Shifman99,DHKM99,HKLM00}. In the Pauli-Villars
scheme it is
 \beq{20}
   <Tr\bar{\psi}\psi>_{\rm QFT}={\rm Const.}\ \frac{1}{\gym^2}
    \Lambda^3e^{-\frac{8\pi^2}{N\gym^2}},
 \eeq
where $\Lambda$ is an UV cut-off of QFT. In addition for large
$\rho$ the gauge coupling is approximately
 \beq{21}
   \frac{4\pi^2}{N\gym^2}=\rho+\frac{1}{16}\ln{\rho}
      +c+O(\frac{\ln{\rho}}{\rho}).
 \eeq
It implies that the large $\rho$ limit corresponds to weak 't
Hooft coupling limit of the gauge theory and it agrees with
asymptotic freedom of the ${\cal N}=1$ pure SYM. Using
Eq.~(\ref{21}), the gaugino condensation~(\ref{12}) can be
rewritten as
 \beq{22}
   <Tr\bar{\psi}\psi>_{\rm SUGRA}&=&\frac{4\pi^2c\eta}{\lambda N}
    \rho_0^{-3/8}e^{2\rho_0}\frac{1}{\gym^2}
    e^{-\frac{8\pi^2}{N\gym^2}} \nonumber \\
    &=&Zl_0^{-3}\frac{1}{\gym^2}e^{-\frac{8\pi^2}{N\gym^2}},
 \eeq
where we define a new radial variable
$l^{-3}=c\lambda^3\rho^{-3/8}e^{2\rho}$ and take cut-off $l=l_0$.
The constants $Z$ and $l$ are independent of $\alpha'$. Comparing
Eqs.~(\ref{20}) with (\ref{22}) we have
 \beq{23} l_0^{-3}\sim \Lambda^3. \eeq
Now let both the radial and the momentum flow away from their
cut-off point we obtain radial/energy-scale relation for the MN
solution
 \beq{24} \rho^{-3/8}e^{2\rho}\sim \frac{\mu^3}{M^3}, \eeq
where $M$ is a definite energy scale.

In order to calculate the two-point function of ${\cal O}_3$ we
have to solve Eq.~(\ref{11}) explicitly. However
equation~(\ref{11}) does not have an explicit analytic solution.
Actually for our purpose we do not need the full analytic solution
we only need asymptotic solution of fluctuation close to the
boundary (or large $\rho$). Notice that both $\delta k$ and
$e^{\delta g}$ are normalized at the boundary, i.e., $\delta k\sim
e^{\delta g}\sim 1$ for $\rho\to\rho_0$ and $\bar{a}\sim
e^{-2\rho}$. Then up to $O(e^{-2\rho}), s$ is decoupled from
$\delta k$ and $\delta g$. Therefore for $\rho\to\infty$ the field
equation~(\ref{11}) is approximately
 \beq{13}
   \ddot{s}+(2-\frac{1}{2\rho})\dot{s}-(n+\frac{p^2}{\lambda^2}
     -\frac{m}{\rho})s=0.
 \eeq
The asymptotic solution of fluctuation field is
 \beq{18}
  \vphi(x,\rho)&\stackrel{\rho\to\infty}{\llraw}&
  \frac{G(\rho)}{G(\rho_0)}e^{ip\cdot x}    \nonumber \\
  G(\rho)&=&\sum_{i=1}^2\lambda_i\rho^{\frac{a_i-2m}{4(a_i+1)}}
  e^{a_i\rho},
 \eeq
with constants $\lambda_i$  and
 \beq{180}a_1&=&-1+\sqrt{1+n+\frac{p^2}{\lambda^2}},\nonumber \\
    a_2&=&-1-\sqrt{1+n+\frac{p^2}{\lambda^2}}.
 \eeq

In general $\lambda_i$ are not constants but are functions of
$p^2/\lambda^2$\cite{Kras02}. As we will see in eq.~(\ref{19}),
however, $\lambda_i$ do not contribute to two-point Green function
when we take cut-off to infinity. Because of this we will
conveniently treat $\lambda_i$ as constants.

Because the operator ${\cal O}_3$ is protected at quantum
correction we expect fluctuation field $\vphi$ to be (mass)
dimension one. In terms of radial/energy-scale relation~(\ref{24})
it implies
 \beq{181} a_1=\frac{2}{3}+O(\frac{p^2}{\lambda^2}), \hspace{0.3in}
     \frac{a_1-2m}{4(a_1+1)}=-\frac{1}{8}+O(\frac{p^2}{\lambda^2}),
 \eeq
with
 \beq{182}m=\frac{1}{2},\hspace{0.5in}n=\frac{16}{9},\eeq
and $a_2=-\frac{8}{3}+O(p^2/\lambda^2)$. The solution~(\ref{18})
is simplified to
 \beq{183} G(\rho)&=&\lambda_1 \rho^{-\frac{1}{8}}e^{2\rho/3}
   \left(1+\frac{p^2}{\lambda^2}(\frac{3}{10}\rho+\frac{27}{500}
   \ln{\rho}) \right. \nonumber \\
   &+&\left.O(e^{-10\rho/3})+O(p^4/\lambda^4)\right). \eeq
Now let us check contribution from fluctuation fields $\delta k$
and $\delta g$. As discussed previously we should add
$O(e^{-2\rho})$ terms in $G(\rho)$ if we consider their
contribution. These terms are $O(e^{-8\rho/3})$ comparing with the
leading order of $G(\rho)$ and therefor we can consistently ignore
those terms.

Using solution~(\ref{18}) and (\ref{183}) the two-point function
is
\begin{widetext}
 \beq{19} <{\cal
  O}_3(p){\cal O}_3(q)>&=&{\rm const.}+\delta^4(p+q)\frac{3c\eta}
   {20\lambda^2}\left(\frac{p^2}{\lambda^2}+O(p^4/\lambda^4)\right)
   (1+\frac{9}{50\rho_0})\rho_0^{-1/2}e^{2\rho_0}
    +O(e^{-2\rho_0/3})\nonumber \\
    &=&{\rm const.}+\delta^4(p+q)p^2\frac{3}{20}Z\frac{\Lambda^3}{\lambda^3}
       [1+\frac{9}{50}(\ln{\frac{\Lambda^2}{M^2}})^{-1}]
       +O(p^4/\lambda^4)+O(1/\Lambda).
 \eeq
\end{widetext}
From this result we see that there are no finite terms in
two-point function of ${\cal O}_3$. At low energy $O(p^4)$ terms
are suppressed by $\alpha'$. Then only $p^2$ terms survive. The
above result is important: it implies that gauginos is a dynamical
composite field, like quarks composed by mesons in QCD.

\subsection{Further discussions}
\label{sec2.4}

Using Eqs.~(\ref{21}) and (\ref{24}) we obtain $\beta$-function of
pure $\net$ SYM
 \beq{25}
   \beta(\gym)=-\frac{3N}{16\pi^2}\gym^3
    [1+\frac{N\gym^2}{16\pi^2}+O(\gym^4)]. \eeq
On the other hand the $\beta$-functions obtained by field theory
method (NSVZ)~\cite{NSVZ} is
\begin{eqnarray}\label{26}
 \beta(\gym)_{\rm NSVZ}=-\frac{3N}{16\pi^2}\gym^3
   [1-\frac{N\gym^2}{8\pi^2}]^{-1}.
\end{eqnarray}

The leading order term of the expression~(\ref{25}) ($\gym^3$)
precisely agree with result of field theory. The sub-leading order
terms, however, does not match with the result of NSVZ. This is a
puzzle of our results. We will see that similar results also
appear for the KST solution. In section~\ref{sec4} we will discuss
several possible solutions for this puzzle.

Another important issue shown in this section is how to establish
field/operator correspondence for the MN solution. For the sake of
convenience let us consider scalar field as an example: All scalar
fields in 5-d bulk are distinguished by two eigenvalues of the
spherical harmonic functions of the transverse space $S^2\times
S^3$ in the UV. It is different from AdS/CFT correspondence that
only one eigenvalue generate ``mass'' of bulk field here we need
both eigenvalues to determine the scaling behavior of bulk field
in the UV. This scaling behavior associates with (mass) dimension
of the dual operator of the boundary quantum field theory.
Therefore field/operator correspondence can be unambiguously
established via correspondence between (mass) dimension of
operator and eigenvalues of the spherical harmonic functions of
the transverse space $S^2\times S^3$.

\section{Green Functions and Radial/Energy-Scale Relation for the
KST Solution}
\label{sec3}

\subsection{The KST solution}

The KST solution is proposed to be dual of $\net$ pure SYM with
gauge group $SU(N+M)\times SU(M)$. It is realized via placing $M$
D3-branes and $N$ fractional D3-branes on the
conifold~\cite{KW99}. If $M$ is a multiple of $N$ then this theory
flows to $SU(N)$ in the IR via a chain of duality cascade which
reduces the rank of the gauge group by $N$ units at each cascade
jump. Thus at the end of the duality cascade the gauge theory is
effectively the $\net$ $SU(N)$ pure SYM. It was shown
in~\cite{KS00} that in order to remove the naked singularity found
in~\cite{KT00} the conifold have to be replaced by the deformed
one. The 10-d metric in the string frame takes ``D-brane'' form:
 \beq{29}
   ds_{10}^2=h^{-1/2}(\tau)dx_ndx_n+h^{1/2}(\tau)ds_6^2, \eeq
where $\tau$ is the radial parameter of the transverse space.
$ds_6^2$ is the metric of the deformed conifold\cite{DC},
 \beq{30}
  ds_6^2&=&\frac{1}{2}\ep^{4/3}K(\tau)
  \left[\frac{1}{3K^3(\tau)}(d\tau^2+(g^5)^2)
  \right.\nonumber \\  &&\left.
   +\cosh^2{(\frac{\tau}{2})}\sum_{i=3}^4(g^i)^2
   +\sinh^2{(\frac{\tau}{2})}\sum_{i=1}^2(g^i)^2
 \right] \eeq
where $\ep$ is a parameter with length dimension $3/2$, $K(\tau)$
is given by
 \beq{31}
 K(\tau)=\frac{(\sinh{(2\tau)}-2\tau)^{1/3}}{2^{1/3}\sinh{\tau}},
 \eeq
and the 1-form basis $g^i$ is defined as follows
 \beq{32}
 g^1&=&-\frac{1}{\sqrt{2}}(\sin{\theta}d\phi_1+\cos{\psi}
   \sin{\theta_2}d\phi_2-\sin{\psi}d\theta_2), \nonumber \\
 g^2&=&\frac{1}{\sqrt{2}}(d\theta_1-\sin{\psi}\sin{\theta_2}
   d\phi_2-\cos{\psi}d\theta_2), \nonumber \\
 g^3&=&-\frac{1}{\sqrt{2}}(\sin{\theta}d\phi_1-\cos{\psi}
   \sin{\theta_2}d\phi_2+\sin{\psi}d\theta_2), \nonumber \\
 g^4&=&\frac{1}{\sqrt{2}}(d\theta_1+\sin{\psi}\sin{\theta_2}
   d\phi_2+\cos{\psi}d\theta_2), \nonumber \\
 g^5&=&d\psi+\cos{\theta_1}d\phi_1+\cos{\theta_2}d\phi_2.
 \eeq
For the KST solution the type IIB SUGRA fields NS-NS 2-form $B-2$,
R-R 2-form $C_2$ and the dilaton $\Phi$ are excited. The simplest
ansatz for the 2-form fields is
 \beq{33} B_2&=&\frac{g_sN\alpha'}{2}[f(\tau)g^1\wedge g^2
    +\kappa(\tau)g^3\wedge g^4],  \\
    F_3&=&\frac{N\alpha'}{2}\{g^5\wedge g^3\wedge g^4+
     d[F(\tau)(g^1\wedge g^3+g^2\wedge g^4)]\}, \nonumber
 \eeq
with boundary condition $F(0)=0$ and $F(\infty)=1/2$. The
self-dual 5-form field strength may be decomposed as
$\td{F}_5={\cal F}_5+*{\cal F}$,
 \beq{34}{\cal F}_5&=&B_2\wedge F_3=\frac{g_sN\alpha'^2}{4}
  l(\tau)g^1\wedge g^2\wedge g^3\wedge g^4\wedge g^5,
      \nonumber \\
  l(\tau)&=&f(1-F)+\kappa F.
 \eeq
Writing the dilation field as
 \beq{35}e^{\Phi}=e^{\Phi_0+\vphi}=g_se^{\vphi}, \eeq
and define
 \beq{36} \alpha=4(g_sN\alpha')^2\ep^{-8/3}, \eeq
the type IIB SUGRA field equations can be rewritten as follows
\begin{widetext}
 \beq{37}
   &&2h\frac{d}{d\tau}(e^{\vphi}h^{-1}F')+e^{\vphi}(1-F)
   \tanh^2{(\frac{\tau}{2})}-e^{\vphi}F\coth^2{(\frac{\tau}{2})}
   =\alpha(\kappa-f)\frac{l}{K^2h\sinh^2{\tau}}, \nonumber \\
   &&h\frac{d}{d\tau}(e^{\vphi}h^{-1}\coth^2{(\frac{\tau}{2})}f')
    -\frac{1}{2}e^{\vphi}(f-\kappa)=\alpha\frac{l(1-F)}
     {K^2h\sinh^2{\tau}}, \nonumber \\
   &&h\frac{d}{d\tau}(e^{\vphi}h^{-1}\tanh^2{(\frac{\tau}{2})}
    \kappa')+\frac{1}{2}e^{\vphi}(f-\kappa)=\alpha\frac{lF}
     {K^2h\sinh^2{\tau}}, \\
   &&\nabla^2\vphi=\frac{\alpha}{8}\ep^{-4/3}
    \{\frac{(1-F)^2e^{\vphi}-\kappa'^2e^{-\vphi}}
     {\cosh^4{(\frac{\tau}{2})}}
     +\frac{F^2e^{\vphi}-f'^2e^{-\vphi}}{\sinh^4{(\frac{\tau}{2})}}
     +8\frac{F'^2e^{\vphi}-(f-\kappa)^2e^{-\vphi}}{\sinh^2{\tau}}\}
       \nonumber \\
    &&\frac{1}{\sinh^2{\tau}}\frac{d}{d\tau}(h'K^2\sinh^2{\tau})=
      -\frac{\alpha}{8}\{\frac{(1-F)^2e^{\vphi}-\kappa'^2e^{-\vphi}}
     {\cosh^4{(\frac{\tau}{2})}}
     +\frac{F^2e^{\vphi}-f'^2e^{-\vphi}}{\sinh^4{(\frac{\tau}{2})}}
     +8\frac{F'^2e^{\vphi}-(f-\kappa)^2e^{-\vphi}}{\sinh^2{\tau}}\}.
         \nonumber
  \eeq
\end{widetext}
Setting $\vphi=0$ we get the KST solution:
 \beq{38} F(\tau)&=&\frac{\sinh{\tau}-\tau}{2\sinh{\tau}},
     \nonumber \\
   f(\tau)&=&\frac{\tau\coth{\tau}-1}{2\sinh{\tau}}(\cosh{\tau}-1),
      \nonumber \\
   \kappa(\tau)&=&\frac{\tau\coth{\tau}-1}{2\sinh{\tau}}
     (\cosh{\tau}+1), \\
   h(\tau)&=&\frac{2^{2/3}}{4}\alpha\int_\tau^\infty dx
     \frac{x\coth{x}-1}{\sinh^2{x}}(\sinh{(2x)}-2x)^{1/3}.
     \nonumber
 \eeq
It should be pointed out that $f$ and $\kappa$ can be shifted by
the same constant. It yields the singular behavior of the metric
for small $\tau$\cite{KS00}.

\subsection{Green functions}

Since NS-NS 2-form is turned on in the KST solution the gauge
coupling of dual SYM is
 \beq{39}
  \frac{1}{\gym^2}=\frac{1}{4\pi^2\alpha'g_{D3}^2}\int d^2\xi
   \sqrt{\det(G_{ij}+B{ij})}. \eeq
In the present case for D5-branes wrapped on the vanishing $S^2$
one has $G=0$. From Eq.~(\ref{32}) and (\ref{33}) we obtain
 \beq{40}
    \frac{1}{\gym^2}=\frac{N}{4\pi^2}\kappa(\tau).
 \eeq
Therefore we will consider Green functions of the following
operator
 \beq{41}{\cal O}_4(x)=-\frac{1}{4}Tr(F_{\mu\nu}F^{\mu\nu})
   +Tr(\bar{\psi}\Sla{D}\psi). \eeq
In terms of AdS/CFT correspondence it requires us to solve
fluctuation equation of $\kappa$. From Eq.~(\ref{38}) however we
can see that functions $\kappa,\;f,\;F$ and $\vphi$ are not
independent from each other. Then we have to solve simultaneous
equations of all fluctuations $\delta\kappa,\;\delta f,\;\delta F$
and $\delta\vphi$. The fluctuation equations do not admit full
analytic solutions. We again consider near-boundary (or large
$\tau$) solution only. In the following all of our discussions and
calculations should be treated at large $\tau$ limit if we do not
stay otherwise.

Conveniently we assume that the fluctuation $\delta F$ vanishes
(in the following we can see this assumption is a consistent one).
Because $e^{\delta\vphi}$ should be normalized at the boundary,
i.e., $e^{\delta\vphi}\to 1$ for $\tau\to\infty$, the fluctuation
equations can be expanded to the first linear order for large
$\tau$. From the fourth equation of Eq.~(\ref{37}) we get
 \beq{42}\na^2\delta\vphi\sim \alpha\ep^{-4/3}e^{-2\tau}
   [\delta\vphi-(\delta\kappa)'-(\delta f)')+O(e^{-4\tau}].
 \eeq
Then noticing $\delta\kappa$ and $\delta f$ are also normalized at
the boundary we have
 \beq{43} \delta\vphi\stackrel{\tau\to\infty}{\llraw} P(\tau)e^{-2\tau}
   +O(e^{-4\tau}), \eeq
where $P(\tau)$ is a polynomial function of $\tau$. Using
Eq.~(\ref{43}) we find that the left side of fluctuation equation
yielded from the first equation of (\ref{37}) is of order
$e^{-2\tau}$. Meanwhile the right side of this equation is
proportional to
 \beq{44} \frac{1}{\tau}(f\delta f-\kappa\delta\kappa)
   +\frac{1}{\tau^2}(f^2-\kappa^2)\delta h. \eeq
Because $f-\kappa\sim e^{-\tau}$ up to $O(e^{-\tau})$ we may set
$\delta f=\delta\kappa=y(x,\tau)$. Then fluctuation equation
yielded from the last equation of (\ref{37}) reduces to
 \beq{45}\frac{d}{d\tau}[(\delta h)'e^{4\tau/3}]\sim\alpha y'. \eeq
Again, recalling $y'\sim O(1/\tau)$, we have $\delta h\sim
O(e^{-4\tau/3}/\tau)$. Finally, up to $O(e^{-\tau})$, we obtain
the fluctuation equation on $y$ ($\delta\kappa$) as follows
 \beq{46} h\frac{d}{d\tau}(h^{-1}y')+\td{\alpha}\tau
   e^{-2\tau/3}\pa_m\pa^m y=(m^2+\frac{4}{3\tau})y, \eeq
where $\td{\alpha}=2^{-10/3}\ep^{4/3}\alpha$ and mass $m$ is
related to eigenvalues of the Laplace equation on $X^5=T^{1,1}$.

We are considering a (mass) dimension four operator. As in AdS/CFT
correspondence the scaling behavior of $y(x,\tau)$ in the UV
requires $m=0$. Then asymptotic solution of Eq.~(\ref{46}) is
 \beq{47} &&y(x,\tau)\stackrel{\tau\to\infty}{\llraw}
   \frac{G(\tau)}{G(\tau_0)}e^{ip\cdot x}, \nonumber \\
   &&G(\tau)=[\tau-\frac{3}{4}+O(1/\tau)]+\frac{3}{8}\td{\alpha}
   p^2[\tau^3+O(\tau^2)]e^{-2\tau/3} \nonumber \\
   &&\quad\quad\quad\quad+O(p^4e^{-4\tau/3}),
 \eeq
where $\tau_0\to\infty$ is a cut-off.

The 5-d action can be obtained via compactifing type IIB SUGAR on
$(g^1,\cdots,g^5)$. In particular only terms involving derivatives
of $\tau$ contribute to one- and two-point functions:
 \beq{48} S_5=\eta\int d^4xd\tau\
 \frac{1}{\tau}e^{4\tau/3}\kappa'^2+\cdots, \eeq
where $\eta\sim N^2/\alpha'^2$. Then the one-point function of
${\cal O}_4$ is
 \beq{49}<{\cal O}_4>=2\eta\frac{1}{\tau_0}e^{4\tau_0/3}, \eeq
and the two-point function is
 \beq{50}  &&<{\cal O}_4(p){\cal O}_4(q)> \nonumber \\
   &=&{\rm const.}+\frac{3}{4}\eta\td{\alpha}[1+O(1/\tau_0)]
   e^{2\tau_0/3}p^2\delta^4(p+q) \nonumber \\
 &&+\lambda_1\eta\td{\alpha}^2\tau_0^2p^4\delta^4(p+q)
   +O(\alpha'p^6),
 \eeq
where $\lambda_1$ is a constant.

\subsection{Radial/energy relation and anomalous dimension}

For large $\tau$ we may introduce another radial coordinate $r$
via $r=\ep^{2/3}e^{-\tau/3}$\cite{KS00}. Noticing
$\td{\alpha}\propto (g_sN\alpha')^2\ep^{-4/3}$ and
 \beq{51} \frac{1}{\gym^2}=\frac{N}{8\pi^2}(\tau-2),\hspace{0.7in}
        \tau\to\infty, \eeq
the two-point function in Eq.~(\ref{50}) can be rewritten as
 \beq{52} &&<{\cal O}_4(p){\cal O}_4(q)> \nonumber \\
  &\sim&(\gym^2 N)^2\delta^4(p+q)[Z_1\ln{r_0}(\ln{r_0}+C)r_0^{-2}p^2
      \nonumber \\
    &&+Z_2\alpha'\ep^{-8/3}(\ln{r_0})^4p^4], \eeq
where $r_0\to 0$ is a cut-off, $Z_1$, $Z_2$ and $C$ are constants
independent of $\alpha'$. Meanwhile the two-point function is easy
to compute at the leading order by perturbative method of quantum
field theory,
 \beq{53} <{\cal O}_4(p){\cal O}_4(q)> &\sim&
   (\gym^2 N)^2[a_0\Lambda^4+a_1p^2\Lambda^2 \nonumber \\ &&+
     a_2p^4\ln{\Lambda^2}+\cdots]\delta^4(p+q). \eeq
Comparing $p^2$ terms in Eqs.~(\ref{50}) and in (\ref{53}) we
obtain the radial/energy-scale relation
 \beq{54} \ln{\mu/M}=\frac{\tau}{3}+\ln{\tau}+{\rm constant}. \eeq
From Eqs.~(\ref{54}) and (\ref{51}) we have the following
$\beta$-function
 \beq{55}\beta(\gym)=-\frac{3N}{16\pi^2}\gym^3
 [1-\frac{3N}{8\pi^2}\gym^2+O(\gym^4)].
 \eeq

Similar as for MN solution we only recover the leading order term
of NSVZ $\beta$-function. The sub-leading order terms,
unfortunately, is very different from that of NSVZ
$\beta$-function. In the next section we will show that there are
some ambiguities which makes it impossible to exactly determine
the radial/energy scale relation beyond leading order in terms of
the above method.

Another puzzle is related to the $p^4$ term of two-point
functions. First we see that the $p^4$ term yielded from SUGRA
description (Eq.~(\ref{52})) is suppressed by $\alpha'$ but not
suppressed from field theoretical method (Eq.~(\ref{53})). In fact
it indicates that $\ep^{8/3}\sim\alpha'$. For KST solution the
singularity of the conifold is removed through the blowing-up of
the $S^3$ of $T^{1,1}$. The length of $S^3$ is parameterized by
$\ep$. Therefore $\ep^{8/3}\sim\alpha'$ means that the 3-sphere is
very small. Recalling $r_0\sim \Lambda/M$ the $p^4$ term yielded
by SUGRA description is proportional to $(\ln{\Lambda^2/M^2})^4$.
However, the one obtained from perturbative calculation of QFT is
proportional to $\ln{\Lambda^2/M^2}$. In the next section, we will
show that, the Green functions yielded from SUGRA description
represent non-perturbative results of QFT. It must correct the
perturbative results of QFT.

In addition, there is a very interesting result hiding in the
solution~(\ref{47}): The dimensions of ${\cal O}_4$ requires that
the leading order function $G(\tau)$ should be a constant. However
it is not true. From eqs.~(\ref{47}), (\ref{51}) and (\ref{54}) we
see that the scaling behavior of $G(\tau)$ is
 \beq{550}G\sim 1+\frac{3}{\tau_0}\ln{\mu}\simeq
  \mu^{\frac{3\gym^2N}{8\pi^2}}. \eeq
It indicates that ${\cal O}_4$ has anomalous dimension
 \beq{551} \gamma=\frac{3\gym^2N}{8\pi^2}. \eeq
In principle this method can be generalized to calculate anomalous
dimensions of any operators.

\section{Discussion}
\label{sec4}

In the previous two sections we have obtained counter-terms of
some Green functions and $\beta$-function of $\net$ pure SYM from
their SUGRA dual description. However there are some mismatchs
between the results from SUGRA description and those from
perturbative calculations of QFT. In order to explain these
mismatchs we should first ask:
\begin{itemize}
\item Are MN solution or KST solution exactly dual to
$\net$ SYM at large N limit?
\item Is the classical action of SUGRA exactly equal to the generation
function of SYM at large N limit?
\end{itemize}
In the case of SUGRA part with maximum supersymmetry and the gauge
theory part conformally invariant the answer should be
unambiguous. In the case of SUGRA part with less supersymmetry and
the gauge theory part not conformally invariant, however, it is
still very difficult to answer these questions thus far. It
requires more careful studies on this type gauge/gravity duality.
In the conformally invariant case we knew that the property of
weak/strong duality prevents us to directly check AdS/CFT
correspondence. The same difficulty also appears our
investigation. It induces ambiguity when we study the divergence
structure of Green functions.

In order to illustrate this ambiguity, let us assume that the
gauge/gravity duality is exact at large $N$ limit for MN solution
or KST solution. It is unambiguous that the SUGRA description
reproduces non-perturbative results of QFT. In indicates that the
Green functions obtained from SUGRA description should contain all
order terms of perturbative calculation of QFT. Usually the high
orders can be ignored at weak 't Hooft coupling limit. It is not
true for the divergence part because higher orders of 't Hooft
coupling are more divergent than lower order. Let us consider KST
solution as an example. Recalling $\tau_0\sim\ln{\Lambda^2/M^2}$
Eq.~(\ref{51}) indicates $(\gym^2N)\ln{\Lambda^2/M^2}$ fixed. The
perturbative calculation of QFT tells us that the most divergence
part of two-point function of ${\cal O}_4$ has the following form
\begin{widetext}
 \beq{56} <{\cal O}_4(p){\cal O}_4(q)>\ \sim\ \sum_{n=0}^{\infty}
 (\gym^2N)^{(n+2)}(\ln{\frac{\Lambda^2}{p^2}})^n(a_{n}p^2\Lambda^2
   +b_{n}p^4\ln{\frac{\Lambda^2}{p^2}}+\cdots)\delta^4(p+q).
   \eeq
\end{widetext}
This implies contributions of every order are equally important!
Of course it is impossible to sum over all order contributions in
terms of perturbative method of QFT. This is a potential origin of
mismatchs mentioned above. However we see that the structure of
$p^2\Lambda^2$ in two-point Green function of ${\cal O}_4$ is
universal. We then obtain the right leading order result on
$\beta$-function. It implies that the SUGRA solution should
include right one-loop effect of dual SYM at least but the
mismatch between Eq.~(\ref{52}) and (\ref{53}) implies the SUGRA
solution must include many higher loop effects of dual SYM.

In order to further understand the above ambiguity, we should
notice that the divergence terms in Green functions is unphysical,
i.e., it has to be subtracted via renormalization procedure. But
this procedure introduces an extra freedom of energy scale
described by the renormalization group equation. We can in general
assume that $p^2$ terms in two-point functions $G^{(2)}(p)$ of any
operator ${\cal O}$ have the following form,
\begin{eqnarray}\label{57}
 G^{(2)}(p^2,\Lambda)_{\rm QFT}\ &\sim&\ \sum_{n=0}^{\infty}
 (\gym^2N)^{(n+m)}(\ln{\frac{\Lambda^2}{m^2}})^na_{n}p^2\Lambda^2,
  \nonumber \\
 G^{(2)}(p^2,\tau_0)_{\rm SUGRA}\ &\sim&\ Z_1P(\tau_0)e^{a\tau_0}p^2,
\end{eqnarray}
where $\tau$ is related to the radial parameter of SUGRA
configuration, $\tau_0\to\infty$ is a cut-off, and
$P(\tau_0)=\tau_0^l+\cdots$ is a polynomial function of $\tau_0$.

Now let both cut-off $\Lambda$ and $\tau_0$ have small changes,
i.e., $\Lambda\to b\Lambda$, $\tau_0\to\tau_0+c$ with $|b-1|<<1$
and $|c|<<1$. It should be careful that $\gym$ in $G^{(2)}(p)_{\rm
QFT}$ is also affected by a rescaling of $\Lambda$. This effect
can be simply captured via treating $\gym$ as energy scale
dependent, i.e., $\gym=\gym(\nu_0)$, and $\nu_0\to b\nu_0$. When
cut-offs are removed we expect to exactly match the results
obtained from SUGRA and QFT respectively. In other words we should
have
\begin{eqnarray}\label{58}
&&\frac{G^{(2)}(p^2,b\Lambda)_{\rm QFT}}{G^{(2)}(p^2,\Lambda)_{\rm
QFT}} =\frac{G^{(2)}(p^2,\tau_0+c)_{\rm
SUGRA}}{G^{(2)}(p^2,\tau_0)_{\rm SUGRA}} \nonumber \\
&\Longrightarrow&\frac{\sum_{n=0}^{\infty}
 [\gym^2(b\nu_0)N]^{(n+m)}(\ln{\frac{\Lambda^2}{m^2}}+\ln{b^2})^na_n}
 {\sum_{n'=0}^{\infty}
 [\gym^2(\nu_0)N]^{(n'+m)}(\ln{\frac{\Lambda^2}{m^2}})^na_{n'}}b^2
   \nonumber \\
&=&\frac{P(\tau_0+c)}{P(\tau_0)}e^{ac}.
\end{eqnarray}
Using result of renormalization group, we can write
\begin{eqnarray}\label{59}
 \gym^2(b\nu_0)=\gym^2(\nu_0)(1+\beta_0\gym^2(\nu_0)\ln{b^2}
  +O(\gym^4(\nu_0)),
\end{eqnarray}
where $\beta_0$ is defined by $\beta$-function
 $$\beta=\beta_0\gym^3+\cdots.$$
Then the second line of Eq.~(\ref{58}) is rewritten as
\begin{eqnarray}\label{60}
&&b^2(1+wg^2(\nu_0)\ln{b^2}), \nonumber \\
w&=&\frac{\sum_na_n(N/s_0)^n[(n+m)\beta_0+ns_0]}
 {\sum_{n'}a_{n'}(N/s_0)^{n'}}.
\end{eqnarray}
Here we have used the relation
$s_0g^2(\nu_0)\ln{\Lambda^2/m^2}\simeq 1$ and $s_0$ depends on
SUGRA solution. For MN solution, $\rho_0\simeq \ln{\Lambda/m}$,
then Eq.~(\ref{18}) gives $s_0=\frac{N}{8\pi^2}$. For KST
solution, $\tau_0\simeq 3\ln{\Lambda/m}$, then Eq.~(\ref{51})
tells us $s_0=\frac{3N}{16\pi^2}$. Let $b=\mu/m$ we can produce a
freedom of energy scale, and $c$ should be related to radial
parameter of SUGRA solutions (named ``$\tau$''). Then we obtain
the radial/energy-scale relation as follows
\begin{eqnarray}\label{61}
c\simeq \frac{1}{a}(1+w
g^2(\nu_0)-\frac{l}{a\tau_0})\ln{\frac{\mu^2}{m^2}}.
\end{eqnarray}
For MN solution, unfortunately , w is not computable so that we
can not obtain radial/energy-scale relation beyond the leading
order. It just reflects the ambiguity mentioned above.

The ambiguity, however, still exists even though Eq.~(\ref{60}) is
computable (e.g., for KST solution). The difficulty is how to
exactly determine the relation between $c$ and $\tau$. There are
two natural conditions: $c(\tau=\tau_0)=0$ and
$dc/d\tau(\tau\to\infty)=1$, but they are not sufficient to fix
the relation between $c$ and $\tau$ beyond leading order for
finite cut-off. For example, the relation
$c=(1+x/\tau_0)(\tau-\tau_0)$ is allowed, but $x$ is an unknown
constant.

In conclusion: We have discussed field/operator correspondence in
gauge/gravity duality in the MN and the KST background. In terms
of this correspondence the counter-terms of some one- and
two-point Green functions of $\net$ SYM have been studied via
their SUGRA dual solution. Although we can only obtain
asymptotical solution of fluctuation equations it is sufficient to
determine some information of the renormalization group equation
of ${\cal N}=1$ SYM. In particular the leading order behavior of
$\beta$-function and anomalous dimension of ${\cal O}_4$ are
obtained. However some ambiguities appear when we want to obtain
information beyond the leading order. It was shown that the extra
constrains are needed to fix these ambiguities. It indicates that
we need to obtain full analytical solutions instead of an
asymptotical solutions of the fluctuation equations. This suggests
more studies are needed to check whether the MN solution or the
KST solution are exactly dual to $\net$ SYM.

\section*{Acknowledgments}

We thank Prof. D. Martelli, J. X. Lu and Ch. J. Zhu for useful
discussions and comments.

\end{document}